\documentclass[reqno,
]{amsproc}
\usepackage{
amssymb}
\allowdisplaybreaks[4]
\DeclareFontFamily{OT1}{wncyi}{}
\DeclareFontShape{OT1}{wncyi}{m}{it}{
   <5> <6> <7> <8> <9> gen * wncyi
   <10> <10.95> <12> <14.4> <17.28> <20.74> <24.88> wncyi10
  }{}
\DeclareSymbolFont{cyrletters}{OT1}{wncyi}{m}{it}
\DeclareSymbolFontAlphabet{\cyrmath}{cyrletters}
\DeclareMathSymbol{\re}{\cyrmath}{cyrletters}{"03}

\newlength{\SomeLenght}
\newcommand*{\hdotsas}[1]{\settowidth{\SomeLenght}{$\displaystyle#1$}
                          \hbox to\SomeLenght{\dotfill}}

\newcommand{\cprime}{\/{\mathsurround=0pt$'$}}
\newcommand{\pinner}{\mathbin{\mathchoice
   {\hbox{\vrule width0.6em depth0pt height0.4pt
   \vrule width0.4pt depth0pt height0.8ex}}
   {\hbox{\vrule width0.6em depth0pt height0.4pt
   \vrule width0.4pt depth0pt height0.8ex}}
   {\hbox{\kern0.14em
   \vrule width0.48em depth0pt height0.4pt
   \vrule width0.4pt depth0pt height0.6ex\kern0.14em}}
   {\hbox{\kern0.1em
   \vrule width0.39em depth0pt height0.4pt
   \vrule width0.4pt depth0pt height0.5ex\kern0.1em}}}}

\DeclareMathOperator{\sym}{sym}
\DeclareMathOperator{\im}{im}

\newcommand{\BBR}{\mathbb{R}}
\newcommand{\CA}{\mathcal{A}}

\newcommand{\CC}{\mathcal{C}}
\newcommand{\CE}{\mathcal{E}}
\newcommand{\CR}{\mathcal{R}}
\newcommand{\CU}{\mathcal{U}}
\newcommand{\Dr}{\mathrm{D}}

\newcommand{\sbs}{\subset}

\newcommand{\ot}{\otimes}

\newcommand{\al}{\alpha}
\newcommand{\be}{\beta}
\newcommand{\Ga}{\Gamma}
\newcommand{\De}{\Delta}
\newcommand{\La}{\Lambda}

\newcommand{\Om}{\Omega}
\newcommand{\om}{\omega}
\newcommand{\vf}{\varphi}
\newcommand{\si}{\sigma}
\newcommand{\h}{-\hspace{0pt}}

\newcommand{\Ei}{\CE^\infty}
\newcommand{\Ci}{C^\infty}
\newcommand{\Ji}{J^\infty}

\def\ldb{{\rm [\![}}
\def\rdb{{\rm ]\!]}}
\newcommand{\fnij}[2]{\ldb{#1},{#2}\rdb}

\newtheorem{theorem}{Theorem}
\newtheorem{proposition}{Proposition}

\theoremstyle{definition}
\newtheorem{definition}{Definition}

\newtheorem*{example}{Example}

\theoremstyle{remark}
\newtheorem{remark}{Remark}

\begin{document}
\title{Complete integrability of the coupled KdV--mKdV system}
\author{Paul Kersten}
\address{University of Twente,
Faculty of Mathematical Sciences,
P.O.Box 217,
7500 AE Enschede,
The Netherlands}
\email{kersten@math.utwente.nl}
\author{Joseph Krasil{\cprime}shchik}
\thanks{Partially supported by the INTAS grant 96-0793.}
\address{Independent University of Moscow\protect\newline
Correspondence to:
1st Tverskoy-Yamskoy per. 14, Apt. 45, 125047 Moscow, Russia}
\email{josephk@online.ru}
\begin{abstract}
The coupled KdV--mKdV system arises as the classical part of one
of superextensions of the KdV equation. For this system, we prove
its complete integrability, i.e., existence of a recursion operator
and of infinite series of symmetries.
\end{abstract}
\maketitle

\section*{Introduction}
There are several supersymmetric extensions of the classical
Korteweg--de Vries equation (KdV) \cite{KrasKers-book,Krivonos,Mathieu}. One
of them is
of the form (the so-called $N=2$, $A=1$ extension \cite{Kers-RIMS})
\begin{align*}
u_t &= - u_3 + 6 u u_1- 3 \vf\vf_2  - 3 \psi\psi_2 - 3 w w_3
   - 3 w_1 w_2 + 3 u_1 w^2 + 6 u w w_1\\
 & + 6 \psi\vf_1 w - 6 \vf\psi_1 w - 6 \vf\psi w_1,\\
\vf_t& =  - \vf_3  + 3 \vf u_1 + 3 \vf_1 u- 3 \psi_2 w
   - 3 \psi_1 w_1  + 3 \vf_1 w^2  + 6 \vf w w_1,\\
\psi_t&=  - \psi_3 + 3 \psi u_1 + 3 \psi_1 u+ 3 \vf_2 w
   + 3 \vf_1 w_1 + 3 \psi_1 w^2 + 6 \psi w w_1,\\
w_t &= - w_3 + 3 w^2 w_1 +3 u w_1 + 3 u_1 w,
\end{align*}
where $u$ and $w$ are classical (even) independent variables while $\vf$
and $\psi$ are odd ones (here and below the numerical subscript at an
unknown variable denotes it derivative over $x$ of the corresponding
order). Being completely integrable itself, this system
gives rise to an interesting system of even equations
\begin{align}\label{intr:eq:kdv-mkdv}
u_t &= - u_3 + 6 u u_1 - 3 w w_3
   - 3 w_1 w_2 + 3 u_1 w^2 + 6 u w w_1,\nonumber\\
w_t &= - w_3 + 3 w^2 w_1 +3 u w_1 + 3 u_1 w,
\end{align}
which can be considered as a sort coupling between the KdV (with
respect to $u$) and the modified KdV (with respect to $w$) equations.
In fact, setting $w=0$, we obtain
\[
u_t = - u_3 + 6 u u_1,
\]
while for $u=0$ we have
\[
w_t = - w_3 + 3 w^2 w_1.
\]

In what follows, we prove complete integrability, cf.~\cite{Dodd}, of system
\eqref{intr:eq:kdv-mkdv} by establishing existence of infinite series of
symmetries and/or conservation laws. Toward this end we construct a recursion
operator using the techniques of deformation theory introduced in
\cite{K-newinv} and extensively described and exemplified in
\cite{KrasKers-book}.

In the first section of the paper the theoretical background is
introduced. The second section deals with particular computations and
description of basic results.

\section{Geometrical and algebraic background}\label{sec:1}
Here we briefly describe the geometrical theory of partial differential
equations \cite{KV-book,KLV} and algebraic foundations of
computational approach to recursion operators \cite{K-newinv,KrasKers-book}.

Let $\pi\colon E\to M$ be a locally trivial vector bundle and
$\pi_k\colon J^k(\pi)\to M$, $k=0,1,\dots,\infty$, be the bundles of
its $k$-jets. A (nonlinear) partial differential equation (PDE) of
order $k$ is a submanifold $\CE\sbs J^k(\pi)$, $k<\infty$. Its $l$th
prolongation is a subset $\CE^l\sbs J^{k+l}(\pi)$. There exist natural
mappings $\pi_{k+l+1,k+l}\colon\CE^{l+1}\to\CE^l$, and $\CE$ is said to
be formally integrable, if all $\CE^l$ are smooth manifolds while
$\pi_{k+l+1,k+l}$ are smooth fiber bundles. Below, only formally integrable
equations are considered.

The inverse limit $\Ei\sbs\Ji(\pi)$ of the system $\{\CE^l,\pi_{k+l,k+l-1}\}$
is called the infinite prolongation of $\CE$ and we consider the bundle
$\pi_\CE\colon\Ei\to M$. This bundle enjoys the following characteristic
property:
\begin{proposition}[see \cite{KLV}]\label{sec1:prop:CC}
Let $\pi_i\colon E_i\to M$\textup{,} $i=1,2$\textup{,} be two locally
trivial vector bundles and $\De\colon\Ga(\pi_1)\to\Ga(\pi_2)$ be a linear
differential operator acting from sections of $\pi_1$ to sections of $\pi_2$.
Then there exists a unique differential operator $\CC\De\colon
\Ga(\pi_\CE^*(\pi_1))\to\Ga(\pi_\CE^*(\pi_2))$ such that
\begin{equation}\label{sec1:eq:CC}
j_\infty(s)^*\circ\CC\De=\De\circ j_\infty(s)^*
\end{equation}
for any formal solution $s$ of the equation $\CE$. The correspondence
$\De\mapsto\CC\De$ is $C^\infty(\Ei)$-linear and complies with the
composition of differential operators\textup{:}
\begin{equation}\label{sec1:eq:CCcomp}
\CC(\De_2\circ\De_1)=\CC\De_2\circ\CC\De_1,
\end{equation}
where $\De_1\colon\Ga(\pi_1)\to\Ga(\pi_2)$\textup{,} $\De_2\colon
\Ga(\pi_2)\to\Ga(\pi_3)$ are linear differential operators\textup{,}
$\pi_3\colon E_3\to M$ being a third vector bundle.
\end{proposition}

As a corollary of Proposition \ref{sec1:prop:CC}, we get
\begin{proposition}\label{sec1:prop:CartConn}
The bundle $\pi_\CE$ possesses a natural flat connection.
\end{proposition}
\begin{proof}
It suffices to take the trivial bundle $\mathbf{1}_M\colon
M\times\mathbb{R}\to M$ for the bundles $\pi_1$ and $\pi_2$ and
an arbitrary vector field $X$ for the operator $\De$. Flatness is
a consequence of \eqref{sec1:eq:CCcomp}.
\end{proof}

Let us denote by $\Dr(N)$ the $C^\infty(N)$-module of vector fields on
a manifold $N$.
\begin{definition}\label{sec1:def9:CartConn}
The connection $\CC\colon\Dr(M)\to\Dr(\Ei)$ is called the
\emph{Cartan connection} on $\Ei$.
\end{definition}

Denote by $\CC\Dr(\Ei)\sbs\Dr(\Ei)$ the horizontal distribution on $\Ei$
with respect to the Cartan connection (the \emph{Cartan distribution}) and
by $\Dr_\CC(\Ei)$ the normalizer of $\CC\Dr(\Ei)$ in $\Dr(\Ei)$. Then, since
$\CC$ is flat, $\CC\Dr(\Ei)$ is integrable in a formal Frobenius sense and
thus $\CC\Dr(\Ei)$ is an ideal in $\Dr_\CC(\Ei)$.
\begin{definition}\label{sec1:def9:symm}
The quotient Lie algebra $\sym\CE=\Dr_\CC(\Ei)/\CC\Dr(\Ei)$ is called the
\emph{algebra of higher symmetries} of the equation $\CE$.
\end{definition}

The Cartan connection in $\pi_\CE$ determines the splitting
\begin{equation}\label{sec1:eq:splitfields}
\Dr(\Ei)=\Dr^v(\Ei)\oplus\CC\Dr(\Ei),
\end{equation}
where $\Dr^v(\Ei)$ denotes the module of
$\pi_\CE$-vertical vector fields, and for any coset $S\in\sym\CE$ there
exists a unique vertical representative. In its turn, such a representative
is uniquely determined by a section $\vf\in\Ga(\pi_\CE^*(\pi))$ (generating
section) satisfying the defining equation
\begin{equation}\label{sec1:eq:ellsymm}
\ell_\CE\vf=0
\end{equation}
and vice versa. Here $\ell_\CE$ is the universal linearization operator
for $\CE$ restricted to $\Ei$. Due to this fact, we shall identify the
solutions of \eqref{sec1:eq:ellsymm} with higher symmetries of $\CE$.

The connection form $U_\CE$ (the \emph{structural element} of $\CE$) of the
Cartan connection $\CC$ is an element of the module $\Dr^v(\La^1(\Ei))$ of
$\La^1(\Ei)$-valued vertical derivations. Thus, we can introduce an operator
$\partial_\CE\colon\Dr^v(\La^i(\Ei))\to\Dr^v(\La^{i+1}(\Ei))$ defined by
\[
\partial_\CE\Om=\fnij{U_\CE}{\Om},\quad\Om\in\Dr^v(\La^i(\Ei)),
\]
where $\fnij{\cdot}{\cdot}$ is the Fr\"olicher--Nijenhuis bracket. Since
the Cartan connection is flat, one has $\fnij{U_\CE}{U_\CE}=0$, from where
it follows that $\partial_\CE\circ\partial_\CE=0$. Thus we obtain a complex
$(\Dr^v(\La^i(\Ei)),\partial_\CE)$, whose cohomology is called the
\emph{$\CC$-cohomology} of $\CE$ and is denoted by $H_\CC^\bullet(\CE)$.
It is easy to see that $H_\CC^0(\CE)=\sym\CE$, while $H_\CC^1(\CE)$, by
standard reasons, is identified with classes of nontrivial infinitesimal
deformations of $U_\CE$ (or, which is the same, of the equation structure).

If $\Om$ and $\Theta$ are elements of $\Dr^v(\La^i(\Ei))$ and
$\Dr^v(\La^j(\Ei))$ respectively, their \emph{contraction} $\Om\pinner
\Theta$ is defined as an element of $(\Dr^v(\La^{i+j-1}(\Ei))$ This
operation is inherited by the $\CC$-cohomology groups. In particular,
if $\vf\in\sym\CE$ and $\CR\in H_\CC^1(\CE)$, then $\vf\pinner\CR=\CR\vf$ is
a symmetry again. In other words, the module $H_\CC^1(\CE)$ acts on the Lie
algebra of higher symmetries.

Let $\CC\La^1(\Ei)\sbs\La^1(\Ei)$ be the submodule of one-forms on $\Ei$
vanishing on the Cartan distribution. Then one has the direct sum
decomposition
\[
\La^1(\Ei)=\CC\La^1(\Ei)\oplus\La_h^1(\Ei)
\]
dual to \eqref{sec1:eq:splitfields}, where $\La_h^1(\Ei)$ is the submodule
of \emph{horizontal} forms. This splitting is also inherited by
$H_\CC^1(\Ei)$ and an element $\CR\in H_\CC^1(\Ei)$ acts nontrivially on
$\sym\CE$ if only it corresponds to a derivation from $\Dr^v(\CC\La^1(\Ei))$.
Moreover, it can be shown that $\im\partial_\CE\cap\Dr^v(\CC\La^1(\Ei))=0$
and consequently nontrivial actions can be found by solving the equation
\begin{equation}\label{sec1:eq:nontriv}
\partial_\CE(\CR)=0.
\end{equation}
Solutions of \eqref{sec1:eq:nontriv} are called \emph{recursion operators}
for symmetries. Any recursion operator $\CR$ is uniquely determined by an
element $\om_\CR\in\CC\La^1(\Ei)\ot\Ga(\pi_\CE(\pi))$ satisfying the
defining equation
\begin{equation}\label{sec1:eq:ellrec}
\ell_\CE^{[1]}(\om_\CR)=0,
\end{equation}
where $\ell_\CE^{[1]}$ is the extension of the operator $\ell_\CE$ to the
module $\CC\La^1(\Ei)\ot\Ga(\pi_\CE(\pi))$. If $\vf\in\Ga(\pi_\CE^*(\pi))$
is a symmetry and $S_\vf\in\Dr^v(\Ei)$ is the corresponding vertical
vector field, then the action of $\CR$ on $\vf$ is given by
\begin{equation}\label{sec1:eq:action}
\CR\vf=S_\vf\pinner\om_\CR.
\end{equation}

\subsection*{Local coordinates}
Let $\CU\sbs M$ be a coordinate neighborhood in $M$ such that the bundle
$\pi$ trivializes over $\CU$, $x_1,\dots,x_n$ be local coordinates in
$\CU$ and $u^1,\dots,u^m$ be coordinates along the fiber in a given
trivialization. Then the \emph{adapted coordinates} in $\pi_\infty^{-1}(\CU)
\sbs\Ji(\pi)$ are given by the functions $u_\si^j$, $j=1,\dots,m$, $\si=
i_1\dots i_k$, $1\le i_\al\le n$, uniquely defined by
\[
j_\infty(f)^*(u_\si^j)=\frac{\partial^{|\si|}f^j}{\partial x_{i_1}\dots
\partial x_{i_k}}
\]
for any local section $f=(f^1,\dots,f^m)\in\Ga(\left.\pi\right|_\CU)$. Then
the Cartan connection in $\Ji(\pi)$ is given by
\begin{equation}\label{sec1:eq:totder}
\CC\left(\frac{\partial}{\partial x_i}\right)
=D_i=\frac{\partial}{\partial x_i}+\sum_{j,\si}u_{\si i}^j
\frac{\partial}{\partial u_\si^j},
\end{equation}
where $D_i$ are the so-called \emph{total derivatives}. The structural
element in this case is given by the formula
\begin{equation}\label{sec1:eq:Upi}
U_\pi=\sum_{j,\si}\om_\si^j\ot\frac{\partial}{\partial u_\si^j},
\end{equation}
where $\om_\si^j=du_\si^j-\sum_iu_{\si i}^j\,dx_i$ are the \emph{Cartan
forms} constituting a basis in the module $\CC\La^1(\Ji(\pi))$. The forms
$\om_\si^j$ may be rewritten as $\om_\si^j=d_\CC u_\si^j$, where $d_\CC$
is the \emph{Cartan differential} acting on $f\in C^\infty(\Ji(\pi))$ by
\[
d_\CC f=\sum\frac{\partial f}{\partial u_\si^j}\om_\si^j.
\]
This differential restricts to any submanifold $\Ei\sbs\Ji(\pi)$ and
for infinite prolongations \eqref{sec1:eq:Upi} transforms to
\begin{equation}\label{sec1:eq:UE}
U_\CE=\sum_I d_\CC u_I\ot\frac{\partial}{\partial u_I},
\end{equation}
where $\{u_I\}$ spans the set of \emph{internal coordinates} in $\Ei$.

If an equation $\CE\sbs J^k(\pi)$ is given by the system of equalities
\[
\begin{cases}
F^1(x_1,\dots,x_n,\dots,u_\si^j,\dots)=0,\\
\hdotsas{F^1(x_1,\dots,x_n,\dots,u_\si^j,\dots)=0}\\
F^r(x_1,\dots,x_n,\dots,u_\si^j,\dots)=0,
\end{cases}
\]
then its infinite prolongation is defined by
\[
D_\si F^\al=0,\quad\al=1,\dots,r,\ 0\le|\si|<\infty,
\]
where $D_\si=D_{i_1}\circ\dots\circ D_{i_k}$ for $\si=i_1\dots i_k$. Then
the universal linearization operator corresponding to this system is of
the form
\begin{equation}\label{sec1:eq:lincoord}
\ell_F=\left\Vert\sum_\si\frac{\partial F^\al}{\partial u_\si^j}D_\si
\right\Vert.
\end{equation}
The operator $\ell_\CE$ is obtained from \eqref{sec1:eq:lincoord} by
rewriting it in internal coordinates.

\begin{example}[Evolutionary $1+1$ equations]
Consider a system $\CE$ of evolutionary equations
\begin{equation}\label{sec1:eq:evol}
\begin{cases}
u_t^1=F^1(x,t,u^1,\dots,u_k^m),\\
\hdotsas{u_t^1=F^1(x,t,u^1,\dots,u_k^m)}\\
u_t^m=F^r(x,t,u^1,\dots,u_k^m),
\end{cases}
\end{equation}
where $x=x_1$, $t=x_2$ and $u_l^j=\partial^lu^j/\partial x^l$. Then the
functions $x,t,\dots,u_l^j,\dotsb$ can be taken for internal coordinates
on $\Ei$. The total derivatives are written down as
\begin{align*}
D_x=\frac{\partial}{\partial x}+\sum_{l,j}u_{l+1}^j
\frac{\partial}{\partial u_l^j},\quad
D_t=\frac{\partial}{\partial t}+\sum_{l,j}
D_x^l(F^j)\frac{\partial}{\partial u_l^j}
\end{align*}
in these coordinates. The Cartan forms on $\Ei$ are
\[
\om_l^j=du_l^j-u_{l+1}^j\,dx-D_x^l(F^j)\,dt,
\]
while the structural element for $\CE$ is given by
\begin{equation}\label{sec1:eq:struc-evol}
U_\CE=\sum_{l,j}\om_l^j\ot\frac{\partial}{\partial u_l^j}.
\end{equation}

The restriction $\ell_\CE$ of the universal
linearization operator to $\Ei$ is of the form
\[
\ell_\CE=\left\Vert\sum_{l,j}
\frac{\partial F^\al}{\partial u_l^j}D_x^l-\mathbf{E}D_t\right\Vert,
\]
where $\mathbf{E}$ is the identity matrix. So, a vector function $\vf=
(\vf^1,\dots,\vf^m)$, $\vf^j=\vf^j(x,t,\dots, u_l^j,\dots)$, is a symmetry
of $\CE$ if and only if
\begin{equation}\label{sec1:eq:symevol}
\sum_{l,j}\frac{\partial F^\al}{\partial u_l^j}D_x^l\vf^j=D_t\vf^\al
\end{equation}
for all $\al=1,\dots,m$. The corresponding vector field is given by the
formula
\[
\re_\vf=\sum_{l,j}D_x^l(\vf^j)\frac{\partial}{\partial u_l^j}.
\]

In a similar way, recursion operators are determined by vector-valued forms
$\om_\CR=(\om_\CR^1,\dots,\om_\CR^m)$, where $\om_\CR^j=
\sum_{l,\al}\psi_\al^{jl}\om_l^\al$, $\psi_\al^{jl}\in C^\infty(\Ei)$,
satisfying the equations
\begin{equation}\label{sec1:eq:recevol}
\sum_{l,j}\frac{\partial F^\al}{\partial u_l^j}D_x^l\om_\CR^j=D_t\om_\CR^\al.
\end{equation}
To compute the left and right sides of \eqref{sec1:eq:recevol}, it suffices
to note that
\[
D_x\om_l^j=\om_{l+1}^j,\quad D_t\om_l^j=D_x^ld_\CC F^j=
D_x^l\sum_{\al,be}\frac{\partial F^j}{\partial u_\beta^\al}\om_\beta^\al.
\]
\end{example}

\begin{remark}\label{sec1:rem:action}
Let $\om=(\om^1,\dots,\om^m)$, $\om^j=\sum_{l,\al}\psi_\al^{jl}\om_l^\al$,
be a vector-valued Cartan form on $\Ei$. Then for any vector-valued
function $\vf=(\vf^1,\dots,\vf^m)$ on $\Ei$ the action $\CR_\om\colon\vf
\mapsto\CR_\om\vf$ is defined by $\CR_\om\vf=\re_\vf\pinner\om$. In local
coordinates, this action is expressed by the formula
\begin{equation}\label{sec1:eq:action1}
(\CR_\om\vf)^j=\sum_{l,\al}\psi_\al^{jl}D_x^l(\vf^\al).
\end{equation}
Operators of this type (i.e., expressed in terms of total derivatives)
are called \emph{$\CC$\h differential} (or \emph{total differential})
operators.
\end{remark}

\begin{remark}\label{sec1:rem:adj}
As it was mentioned above, operators of the form \eqref{sec1:eq:action1},
provided $\om$ satisfies \eqref{sec1:eq:recevol}, take symmetries of
the equation at hand to symmetries of the same equation. In other words,
one has $\CR_\om\colon\ker\ell_\CE\to\ker\ell_\CE$. Under not very
restrictive conditions on the equation $\CE$, this is equivalent to the
operator equality
\[
\ell_\CE\circ\CR_\om=\CA\circ\ell_\CE,
\]
where $\CA$ is a $\CC$-differential operator. Taking formally adjoint,
one obtains
\[
\CR_\om^*\circ\ell_\CE^*=\ell_\CE^*\circ\CA^*,
\]
which means, that $\CA^*$ is a recursion operator for generating
functions of conservation laws \cite{AMV-local}.
\end{remark}

\subsection*{Nonlocal setting}
In practice, when solving \eqref{sec1:eq:ellrec}, one usually finds no
nontrivial solutions, though the equation $\CE$ may possess recursion
operators.
\begin{example}[The Burgers equation]\label{sec1:ex:burg}
Consider the equation
\[
u_t=u_{xx}+uu_x.
\]
It is known to possess a recursion operator of the form
\begin{equation}\label{sec1:eq:Burrec}
\CR=D_x+\frac12u_0+\frac12u_1D_x^{-1}.
\end{equation}
Nevertheless, the only solution of the equation
\begin{equation}\label{sec1:eq:Burellrec}
\ell_\CE^{[1]}\om\equiv (D_x^2+u_0D_x+u_1-D_t)\om=0
\end{equation}
for $\om=\psi_0\om_0+\dots\psi_k\om_k$ is $\al\om_0$, $\al\in\BBR$, which
provides the trivial action $\CR_\om\colon\vf\mapsto\al\vf$.

To resolve this apparent contradiction, let us extend the algebra $\Ci(\Ei)$
with an additional element $u_{-1}$ and set
\begin{align}\label{sec1:eq:Burcov}
D_xu_{-1}&=u_0,\nonumber\\
D_tu_{-1}&=u_1+\frac12u_0^2,\nonumber\\
d_\CC u_{-1}&\equiv\om_{-1}=du_{-1}-u_0\,dx-\left(u_1+\frac12u_0^2\right)dt.
\end{align}
Then, solving \eqref{sec1:eq:Burellrec} for $\om=\psi_{-1}\om_{-1}
+\psi_0\om_0+\dots\psi_k\om_k$, we obtain a two\h parametric solution
\[
\om=\al\om_0+\beta\Om,\quad\Om=\om_1+\frac12u_0\om_0+\frac12u_1\om_{-1}.
\]
Then the action $\vf\mapsto\CR_\Om\vf=\re_\vf\pinner\Om$ coincides exactly
with \eqref{sec1:eq:Burrec}.
\end{example}

This example reflects a general scheme of computations which arises in
a lot of applications \cite{KrasKers-book} and is used in Section
\ref{sec:2}. Namely, in search of recursion operators for symmetries
we extend the algebra $C^\infty(\Ei)$ with a new set of variables
$w^1,\dots,w^r,\dots$ (so-called \emph{nonlocal variables}) and respectively
extend the total derivatives to vector fields
\begin{equation}\label{sec1:eq:totder-ev}
\bar{D}_i=D_i+\sum_\al X_i^\al\frac{\partial}{\partial w^\al},\quad
i=1,\dots,n,
\end{equation}
in such a way that
\begin{equation}\label{sec1:eq:totder1}
[\bar{D}_i,\bar{D}_j]\equiv[D_i,X_j]+[X_i,D_j]+[X_i,X_j]=0,
\end{equation}
where $X_l=\sum_\al X_l^\al{\partial}/{\partial w^\al}$.

Having a solution $X_1,\dots,X_n$ of \eqref{sec1:eq:totder-ev}, we obtain an
integrable distribution on the space $\Ei\times\BBR^N$, where $N$ is the
number of nonlocal variables (the case $N=\infty$ is included). The
projection $\tau\colon\overline{\Ei}=\Ei\times\BBR^N\to\Ei$ is called a
\emph{covering} over $\CE$ and $N$ is called its \emph{dimension} (for an
invariant geometrical definition see \cite{KV-trends}). Similar to the local
case, we define the Lie algebra $\sym_\tau\CE=\Dr_\CC(\overline{\Ei})/
\CC\Dr(\overline{\Ei})$ of \emph{nonlocal} $\tau$\h symmetries.

We introduce \emph{Cartan forms}
\[
\theta^j=dw^j-\sum_{i=1}^n X_i^j\,dx_i,\quad j=1,\dots,N,
\]
corresponding to nonlocal variables on $\overline{\Ei}$. The module of all
Cartan forms on $\overline{\Ei}$ is denoted by $\CC\La^1(\overline{\Ei})$. We
also extend the universal linearization operator $\ell_\CE$ to
$\overline{\Ei}$ just by changing the total derivatives $D_i$ to $\bar{D}_i$.
Let us now consider two equations, associated to this extension:
\[
\bar{\ell}_\CE\vf=0\ \text{ and }\ \bar{\ell}_\CE^{\,[1]}\Om=0,
\]
where $\vf\in\Ga((\pi_\CE\circ\tau)^*\pi)$ and $\Om\in
\Ga((\pi_\CE\circ\tau)^*\pi)\ot\CC\La^1(\overline{\Ei})$. Solutions of the
first equation are called $\tau$\h\emph{shadows} of nonlocal symmetries,
while solutions of the second one are said to be $\tau$\h\emph{shadows} of
recursion operators in the covering $\tau$. The following result establishes
relations of shadows to symmetries and recursion operators:
\begin{theorem}[see \cite{KrasKers-book,KV-trends}]
Let $\tau\colon\overline{\Ei}\to\Ei$ be a covering. Then\textup{:}
\begin{enumerate}
\item
If $\vf$ is a $\tau$-shadow\textup{,} then there exists a covering
\[
\bar{\tau}\colon\overline{\overline{\Ei}}\to\overline{\Ei}
\xrightarrow{\tau}\Ei
\]
such that $\vf$ reconstructs up to a nonlocal $\bar{\tau}$\h symmetry.
\item
If $\vf$ is a nonlocal $\tau$\h symmetry and $\CR$ is a recursion operator
shadow\textup{,} then $\CR\vf$ is a symmetry shadow.
\end{enumerate}
\end{theorem}
\begin{remark}\label{sec1:rem:horcohom}
Among all one-dimensional coverings over a given equation there exists a
special class consisting of those ones, for which the fields $X_1,\dots,
X_n$ in \eqref{sec1:eq:totder1} are independent of nonlocal variables
(so-called \emph{abelian coverings}). To any
such a covering, one can put into correspondence a differential form on
$\Ei$:
\[
\om_\tau=\sum_{l=1}^nX_ldx_l.
\]
This form is closed with respect to the so-called \emph{horizontal
differential} $d_h=\CC d$ (cf.~Proposition \ref{sec1:prop:CC}). Vice versa,
to any such a form there corresponds
a covering of the above mentioned type. Moreover, two closed forms determine
the same class in the cohomology group $H^1(d_h)$ if and only if
the corresponding coverings are equivalent. In particular, if $n=2$, the
group $H^1(d_h)$ coincides with the group of conservation laws of the
equation $\CE$ \cite{AMV-local}. Thus, to construct a covering under
consideration is the same as to find a conservation law. This fact is used
in the computations below.
\end{remark}

\section{Basic computations and results}\label{sec:2}
In this section we shall discuss the complete integrability of the KdV--mKdV
system given in \eqref{intr:eq:kdv-mkdv}, i.e.,
\begin{align}\label{sec2:eq:kdv-mkdv}
u_t &= - u_3 + 6 u u_1 - 3 w w_3
   - 3 w_1 w_2 + 3 u_1 w^2 + 6 u w w_1,\nonumber\\
w_t &= - w_3 + 3 w^2 w_1 +3 u w_1 + 3 u_1 w.
\end{align}
In order to demonstrate the complete integrability of this system, we shall
construct the recursion operator for symmetries of this coupled system,
leading to infinite hierarchies  of symmetries and, most probably, of
conservation laws. Due to the very special form of the final results, it
seems that integrability of this system, which looks quite ordinary, has not
been discussed before or elsewhere. In order to do this, we shall discuss
conservation laws in Subsection \ref{sec:2.1} leading to the necessary
nonlocal variables.

In Subsection \ref{sec:2.2} we shall discuss  local and nonlocal symmetries of the system,
while in Subsection \ref{sec:2.3} we construct the recursion operator or
deformation of the equation structure \eqref{sec1:eq:struc-evol}.

\subsection{Conservation laws and nonlocal variables}\label{sec:2.1}
Here we shall construct conservation laws for \eqref{sec2:eq:kdv-mkdv}
in order to arrive at an abelian covering of the coupled KdV--mKdV
system as was shown for Burgers equation \eqref{sec1:eq:Burcov}.
So we construct $X=X(x,t,u,\ldots,w\ldots),T=T(x,t,u,\ldots,w\ldots)$
such that
\begin{align}
D_x(T)=D_t(X)
\end{align}
and in a similar way we construct nonlocal conservation laws by the
requirement
\begin{align}
\bar{D}_x(\bar{T})=\bar{D}_t(\bar{X}),
\end{align}
where $\bar{D}_*$ is defined by \eqref{sec1:eq:totder}; moreover $\bar{X}$,
$\bar{T}$ are dependent on local variables $x$, $t$, $u,\dots,w,\dots$ as
well as the already determined nonlocal variables, denoted here by $p_*$ or
$p_{*,*}$, which are associated to the conservation laws $(X,T)$ by the
formal definition
\begin{align*}
D_x(p_{*})&=(p_{*})_x=X,\\
D_t(p_{*})&=(p_{*})_t=T.
\end{align*}
Proceeding in this way, we obtained the following set of nonlocal variables
\begin{equation}\label{kdv-aug}
p_{0,1},\ p_{0,2},\ p_1,\ p_{1,1},\ p_{1,2},\ p_{2,1},\ p_3,\ p_{3,1},\
p_{3,2},\ p_{4,1},\ p_5,
\end{equation}
where their defining equations are given by
\begin{align*}
(p_1)_x& = u,\\
(p_1)_t &= 3 u^2 + 3 u w^2 - u_2 - 3 w w_2,\\
(p_{0,1})_x &= w,\\
(p_{0,1})_t &= 3 u w + w^3 - w_2,\\
(p_{0,2})_x &= p_1,\\
(p_{0,2})_t &=  - 6 p_3 - u_1,\\
(p_{1,1})_x &= \cos(2 p_{0,1}) p_1 w + \sin(2 p_{0,1}) w^2,  \\
(p_{1,1})_t &= \cos(2 p_{0,1}) (3 p_1 u w + p_1 w^3 - p_1 w_2 + u w_1-u_1 w
-w^2 w_1)\\
&+\sin(2 p_{0,1}) (4 u w^2 + w^4 - 2 w w_2 + w_1^2),  \\
(p_{1,2})_x &= \cos(2 p_{0,1}) w^2 - \sin(2 p_{0,1}) p_1 w,  \\
(p_{1,2})_t &= \cos(2 p_{0,1}) (4 u w^2 + w^4 - 2 w w_2 + w_1^2)\\
+&\sin(2 p_{0,1})(-3 p_1 u w - p_1 w^3 + p_1 w_2 - u w_1 + u_1 w + w^2 w_1),\\
(p_{2,1})_x&=(4\cos(2 p_{0,1})p_{1,1}w^2-4\sin(2p_{0,1})p_1 p_{1,1}w
+w (p_1^2 - 2 u + w^2))/2,\\
(p_{2,1})_t &= (4 \cos(2 p_{0,1}) p_{1,1} (4 u w^2 + w^4 - 2 w w_2+w_1^2)\\
&+4\sin(2 p_{0,1})p_{1,1}(-3p_1u w-p_1w^3+p_1w_2- u w_1 + u_1 w + w^2 w_1)\\
& + 3 p_1^2 u w + p_1^2 w^3 - p_1^2 w_2 + 2 p_1 u w_1 - 2 p_1 u_1 w -
2 p_1 w^2 w_1 - 8 u^2 w   \\
&- u w^3+ 2 u w_2 - 2 u_1 w_1 + 2 u_2 w + w^5 + 3 w^2 w_2)/2,  \\
(p_3)_x &= ( - u^2 - u w^2 + w w_2)/2,  \\
(p_3)_t &=(-4 u^3 - 9 u^2 w^2 + 2 u u_2 - 3 u w^4 + 11 u w w_2 - u w_1^2 -
u_1^2 + u_1 w w_1  \\
& + 4 u_2 w^2 + 6 w^3 w_2 + 3 w^2 w_1^2 - w w_4 + w_1 w_3 - w_2^2)/2,  \\
(p_{3,1})_x&=(\cos(2p_{0,1}) w (p_1^3 - 6 p_1 u + 39 p_1 w^2 - 24 p_{1,1}
p_{1,2} w + 12 p_3 + 6 u_1)\\
&+2 \sin(2 p_{0,1}) w (12 p_1 p_{1,1} p_{1,2} + 18 p_1 w_1 + 2 w^3 + 3
w_2)  \\
& + 6p_{1,2} w ( - p_1^2 + 2 u - w^2))/12,  \\
(p_{3,2})_x&=(2\cos(2p_{0,1})w(12 p_1 p_{1,1}p_{1,2}-18 p_1 w_1-2 w^3-3w_2)\\
&+ \sin(2 p_{0,1}) w (p_1^3 - 6 p_1 u + 39 p_1 w^2 + 24 p_{1,1} p_{1,2}w
+ 12 p_3 + 6 u_1)  \\
& + 6 p_{1,1} w( - p_1^2 + 2 u - w^2))/12,  \\
(p_{4,1})_x &= (8 \cos(2 p_{0,1}) w (p_1^3 p_{1,2} + 12 p_1 p_{1,1}^2 p_{1,2}
- 6 p_1 p_{1,2} u + 3 p_1 p_{1,2} w^2  \\
& - 12 p_{1,1} p_{1,2}^2 w + 18 p_{1,1} u w - 4 p_{1,1} w^3 - 6 p_{1,1} w_2
+ 12 p_{1,2} p_3 + 6 p_{1,2} u_1)   \\
&+ 8 \sin(2 p_{0,1}) w (p_1^3 p_{1,1} + 12 p_1 p_{1,1} p_{1,2}^2 - 6
p_1 p_{1,1} u + 3 p_1 p_{1,1} w^2  \\
& + 12 p_{1,1}^2 p_{1,2} w+ 12 p_{1,1} p_3
 + 6 p_{1,1} u_1 - 18 p_{1,2} u w + 4 p_{1,2} w^3 + 6 p_{1,2} w_2)  \\
& + w ( - p_1^4 - 24 p_1^2 p_{1,1}^2 - 24 p_1^2 p_{1,2}^2 + 12 p_1^2 u
- 6 p_1^2 w^2 - 48 p_1 p_3   \\
&- 24 p_1 u_1 + 48 p_{1,1}^2 u - 24 p_{1,1}^2 w^2 + 48 p_{1,2}^2 u
-24 p_{1,2}^2 w^2  \\
& - 60 u^2 + 44 u w^2 + 24 u_2 - 13 w^4 + 6 w w_2))/48,  \\
(p_5)_x &= (12 u^3 + 24 u^2 w^2 - 6 u u_2 + 6 u w^4 - 30 u w w_2
- 3 u_2 w^2 - 8 w^3 w_2 + 6 w w_4)/6.
\end{align*}

In the previous equations, we skipped explicit formulas for $(p_{3,1})_t$,
$(p_{3,2})_t$, $(p_{4,1})_t$, and $(p_5)_t$, because they are too massive,
though quite important for the setting to be well defined and in order to
avoid ambiguities. The reader is referred to the Appendix for them.

It is quite a striking result that functions $\cos(2p_{0,1})$,
$\sin(2p_{0,1})$ appear in the presentation of the conservation laws and
their associated nonlocal variables.

We should note that $p_1$, $p_{0,1}$, $p_3$, $p_5$ arise from \textbf{local
conservation laws} and we shall call $p_1$, $p_{0,1}$, $p_3$, $p_5$
\emph{nonlocalities of first order}.

In a similar way we see that  $p_{0,2}$, $p_{1,1}$, $p_{1,2}$ arise from
\textbf{nonlocal conservation laws}, where their $x$- and $t$-derivatives are
dependent on the first order nonlocalities. For this reason $p_{0,2}$,
$p_{1,1}$, $p_{1,2}$ are called \emph{nonlocalities of second order}.

Proceeding in this way $p_{2,1}$, $p_{3,1}$, $p_{3,2}$, $p_{4,1}$ constitute
\emph{nonlocalities of third  order}.

\subsection{Local and nonlocal symmetries}\label{sec:2.2}
In this section we shall present results for the construction of local and
nonlocal symmetries of system \eqref{sec2:eq:kdv-mkdv}. In order to construct
these symmetries, we consider the system of partial differential equations
obtained by the infinite prolongation of \eqref{sec2:eq:kdv-mkdv} together
with the covering by the nonlocal variables
\[
p_{0,1},\ p_{0,2},\ p_1,\ p_{1,1},\ p_{1,2},\ p_{2,1},\ p_3,\ p_{3,1},\
p_{3,2},\ p_{4,1},\ p_5.
\]
So, in the augmented setting governed by \eqref{sec2:eq:kdv-mkdv}, their
total derivatives and the equations given in Subsection \ref{sec:2.1} we
construct symmetries $Y=(Y^u,Y^w)$ which have to satisfy the symmetry
condition
\[
\bar{\ell}_\CE Y=0.
\]
From this condition we obtained the following symmetries
\[
Y_{0,1},\ Y_{1,1},\ Y_{1,2},\ Y_{1,3},\ Y_{2,1},\ Y_{3,1},\ Y_{3,2},\ Y_{3,3},
\]
where generating functions $Y_{*,*}^u$, $Y_{*,*}^w$ are given as
\begin{align*}
Y_{0,1}^u &= 3t(6 u u_{1} + 6uww_{1} + 3 u_{1} w^2 - u_{3} - 3 w w_{3}
- 3 w_{1} w_{2}) + xu_{1} + 2 u, \\
Y_{0,1}^w &= 3t(3 u w_{1} + 3u_{1} w + 3 w^2 w_{1} - w_{3}) + x w_{1} + w, \\
Y_{1,1}^u &= u_{1}, \\
Y_{1,1}^w &= w_{1}, \\
Y_{1,2}^u &= \cos(2 p_{0,1}) (2 u w - w_{2}) + \sin(2 p_{0,1}) (u_{1}
+ 2 w w_{1}), \\
Y_{1,2}^w &=  - \cos(2 p_{0,1}) u - \sin(2 p_{0,1}) w_{1}, \\
Y_{1,3}^u &= \cos(2 p_{0,1}) (u_{1} + 2 w w_{1})
+ \sin(2 p_{0,1}) ( - 2 u w + w_{2}), \\
Y_{1,3}^w &=  - \cos(2 p_{0,1}) w_{1} + \sin(2 p_{0,1}) u, \\
Y_{2,1}^u &= (2 \cos(2 p_{0,1}) (p_{1,1} u_{1} + 2 p_{1,1} w w_{1} - 2
p_{1,2} u w + p_{1,2} w_{2}) \\
& + 2 \sin(2 p_{0,1}) ( - 2 p_{1,1} u w + p_{1,1} w_{2} - p_{1,2} u_{1}
- 2 p_{1,2} w w_{1})  \\
&+ 2 p_{1} u w -
p_{1} w_{2} + 2 u w_{1} + 3 u_{1} w + 2 w^2 w_{1} - w_{3})/2, \\
Y_{2,1}^w &= (2 \cos(2 p_{0,1}) ( - p_{1,1} w_{1} + p_{1,2} u)
+ 2 \sin(2 p_{0,1}) (p_{1,1} u + p_{1,2} w_{1})\\
&-p_{1} u + u_{1} + w w_{1})/2, \\
Y_{3,1}^u &= (6 u u_{1} + 6 u w w_{1} + 3 u_{1} w^2 - u_{3} - 3 w w_{3}
- 3 w_{1} w_{2})/3, \\
Y_{3,1}^w &= (3 u w_{1} + 3 u_{1} w + 3 w^2 w_{1} - w_{3})/3, \\
Y_{3,2}^u &= (\cos(2 p_{0,1}) ( - 2 p_{1}^2 u w + p_{1}^2 w_{2}
- 4 p_{1} u w_{1} - 6 p_{1} u_{1} w - 4
p_{1} w^2 w_{1} + 2 p_{1} w_{3} \\
& + 8 p_{1,1} p_{1,2} u_{1} + 16 p_{1,1} p_{1,2} w w_{1} - 8 p_{1,2}^2
 u w + 4 p_{1,2}^2 w_{2} - 4 p_{2,1} u_{1} - 8 p_{2,1} w w_{1}  \\
&+ 10 u^2 w + 6 u w^3 - 8 u w_{2} - 14 u_{1} w_{1} - 8 u_{2} w
- 11 w^2 w_{2} - 14 w w_{1}^2 + 2 w_{4}) \\
& + 2 \sin(2 p_{0,1}) ( - 8 p_{1,1} p_{1,2} u w + 4 p_{1,1} p_{1,2} w_{2} - 2 p_{1,2}^2 u_{1} - 4
p_{1,2}^2 w w_{1} + 4 p_{2,1} u w  \\
&- 2 p_{2,1} w_{2} + 6 u u_{1} + 10 u w w_{1} + 3 u_{1} w
^2 - u_{3} + 2 w^3 w_{1} - 3 w w_{3} - 5 w_{1} w_{2}) \\
& + 4 p_{1,2} (2 p_{1} u w - p_{1}
w_{2} + 2 u w_{1} + 3 u_{1} w + 2 w^2 w_{1} - w_{3}))/8, \\
Y_{3,2}^w &= (\cos(2 p_{0,1}) (p_{1}^2 u - 2 p_{1} u_{1} - 2 p_{1} w w_{1} - 8 p_{1,1} p_{1,2} w_{1} + 4
p_{1,2}^2
 u + 4 p_{2,1} w_{1} \\
& - 4 u^2 - 3 u w^2 + 2 u_{2} + 4 w w_{2} + 2 w_{1}^2) \\
& + 2
 \sin(2 p_{0,1}) (4 p_{1,1} p_{1,2} u + 2 p_{1,2}^2 w_{1} - 2 p_{2,1} u - 3 u w_{1} - 3 u_{1} w
- 3 w^2 w_{1} + w_{3}) \\
& + 4 p_{1,2} ( - p_{1} u + u_{1} + w w_{1}))/8, \\
Y_{3,3}^u &= (2 \cos(2 p_{0,1}) (2 p_{1,1}^2 u_{1} + 4 p_{1,1}^2 w w_{1} - 4 p_{2,1} u w + 2 p_{2,1}
w_{2} -6 u u_{1}\\
& - 10 u w w_{1}  - 3 u_{1} w^2 + u_{3} - 2 w^3 w_{1} + 3 w w_{3} + 5 w_{1}
w_{2}) \\
& + \sin(2 p_{0,1}) ( - 2 p_{1}^2 u w + p_{1}^2 w_{2} - 4 p_{1} u w_{1} - 6 p_{1}
u_{1} w - 4 p_{1} w^2 w_{1} + 2 p_{1} w_{3} \\
& - 8 p_{1,1}^2 u w + 4 p_{1,1}^2 w_{2} - 4
p_{2,1} u_{1} - 8 p_{2,1} w w_{1} + 10 u^2 w + 6 u w^3 \\
& - 8 u w_{2} - 14 u_{1} w_{1} -
 8 u_{2} w - 11 w^2 w_{2} - 14 w w_{1}^2 + 2 w_{4}) \\
& + 4 p_{1,1} (2 p_{1} u w - p_{1}
 w_{2} + 2 u w_{1} + 3 u_{1} w + 2 w^2 w_{1} - w_{3}))/8, \\
Y_{3,3}^w &= (2 \cos(2 p_{0,1}) ( - 2 p_{1,1}^2 w_{1} + 2 p_{2,1} u + 3 u w_{1} + 3 u_{1} w + 3 w^2
w_{1} - w_{3}) \\
& + \sin(2 p_{0,1}) (p_{1}^2 u - 2 p_{1} u_{1} - 2 p_{1} w w_{1} + 4 p_{1,1}^2
u + 4 p_{2,1} w_{1} - 4 u^2 \\
& - 3 u w^2 + 2 u_{2} + 4 w w_{2} + 2 w_{1}^2) \\
& + 4p_{1,1} ( - p_{1} u + u_{1} + w w_{1}))/8.
\end{align*}

\subsection{Recursion operator}\label{sec:2.3}
Here we present the recursion operator $\CR$ for symmetries for this case
obtained as a higher symmetry in the Cartan covering of system of equations
\eqref{intr:eq:kdv-mkdv} augmented by equations governing the nonlocal
variables \eqref{kdv-aug}. As explained in the previous section, the
recursion operator is in effect the a deformation of the equation structure
\eqref{sec1:eq:UE}.

As demonstrated there, this deformation is a form-valued vector field (or a
vectorfield-valued  one-form)  and has to satisfy
\begin{align}
\bar{\ell}_\CE^{\,[1]}\CR=0.
\end{align}
In order to arrive at a nontrivial result as was explained for Burgers'
equation too (c.f.~Example \ref{sec1:ex:burg}), we have to introduce
nonlocal variables
\[
p_{0,1},\ p_{0,2},\ p_1,\ p_{1,1},\ p_{1,2},\ p_{2,1},\ p_3,\ p_{3,1},\
p_{3,2},\ p_{4,1},\ p_5
\]
and their associated Cartan contact forms
\[
\om_{p_{0,1}},\ \om_{p_{0,2}},\ \om_{p_1},\ \om_{p_{1,1}},\ \om_{p_{1,2}},\
\om_{p_{2,1}},\ \om_{p_3},\ \om_{p_{3,1}},\ \om_{p_{3,2}},\ \om_{p_{4,1}},\
\om_{p_5}.
\]
The final result, which is dependent on the nonlocal Cartan forms
\[
\om_{p_{0,1}},\ \om_{p_1},\ \om_{p_{1,1}},\ \om_{p_{1,2}},
\]
is given by
\begin{equation}\label{skdv2.-2.:R1}
\CR=R^u \frac{\partial}{\partial u}+R^w \frac{\partial}{\partial w}+\dots,
\end{equation}
where the components $R^u$, $R^w$ are given by
\begin{align}\label{skdv2.-2.R2}
R_u&=\om_{u_2}(-1)+ \om_{u}(4u+w^2)+\om_{w_2}(-2w)+\om_{w_1}(-w_1)+\om_{w}(3uw-2w_2)\nonumber\\
   &+\om_{p_{1,2}}(-\cos(2p_{0,1})(u_1+2ww_1)+\sin(2p_{0,1})(2uw-w_2))\nonumber\\
   &+\om_{p_{1,1}}(\cos(2p_{0,1})(-2uw+w_2)-\sin(2p_{0,1})(u_1+2ww_1))\nonumber\\
   &+\om_{p_{1}}(2u_1+ww_1)+\om_{p_{0,1}}(2p_1uw-p_1w_2+2uw_1+3u_1w+2w^2w_1-w_3),\nonumber\\
R_w&=\om_{w_2}(-1)+ \om_{w}(2u+w^2)+\om_{u}(2w)\nonumber\\
   & +\om_{p_{1,2}}(\cos(2p_{0,1})w_1-\sin(2p_{0,1})u)\nonumber\\
   & +\om_{p_{1,1}}(\cos(2p_{0,1})u+\sin(2p_{0,1})w_1)\nonumber\\
   &+\om_{p_{1}}(w_1)+\om_{p_{0,1}}(-p_1u+u_1+ww_1).
\end{align}

We shall now present this result in a more conventional form which appeals
to expressions using operators of the form $D_x$ and $D_x^{-1}$. In order to
do this, we first split \eqref{skdv2.-2.R2} into the so-called local part and
nonlocal parts, consisting of terms associated to $\om_{u_2}$, $\om_u$,
$\om_{w_2}$, $\om_{w_1}$, $\om_w$ and those associated to $\om_{p_{1,2}}$,
$\om_{p_{1,1}}$, $\om_{p_{1}}$, $\om_{p_{0,1}}$ respectively. The first
part will account for $D_x$ presentation, while the second one accounts for
the $D_x^{-1}$ part.

Due to the action of contraction $\re_\vf\pinner\CR$, the local part is
given by the following matrix operator:
\[
\begin{bmatrix}
-D_x^2+4u+w^2&-2wD_x^2-w_1D_x+3uw-2w_2\\
2w&-D_x^2+2u+w^2
\end{bmatrix}.
\]
The nonlocal part will be split into parts associated to $\om_{p_1}$,
$\om_{p_{0,1}}$ and $\om_{p_{1,2}}$, $\om_{p_{1,1}}$, respectively. The
first one is given as
\[
\begin{bmatrix}
(2u_1+ww_1)D_x^{-1}&(2p_1uw-p_1w_2+2uw_1+3u_1w+2w^2w_1-w_3)D_x^{-1}\\
w_1D_x^{-1}&(-p_1u+u_1+ww_1)D_x^{-1}
\end{bmatrix}.
\]

To deal with the last part, let us introduce the notation:
\begin{align*}
A_1&=\cos(2p_{0,1})(-2uw+w_2)-\sin(2p_{0,1})(u_1+2ww_1),\\
A_2&=\cos(2p_{0,1})u+\sin(2p_{0,1})w_1,\\
B_1&=-\cos(2p_{0,1})(u_1+2ww_1)+\sin(2p_{0,1})(2uw-w_2),\\
B_2&=\cos(2p_{0,1})w_1-\sin(2p_{0,1})u,
\end{align*}
being the coefficients at $\om_{p_{1,1}}$ and $\om_{p_{1,2}}$ in
\eqref{skdv2.-2.R2}.

According to the presentations of $(p_{1,1})_x$ and $(p_{1,2})_x$, i.e.,
\begin{align*}
(p_{1,1})_x &= \cos(2 p_{0,1}) p_1 w + \sin(2 p_{0,1}) w^2\\
(p_{1,2})_x &= \cos(2 p_{0,1}) w^2 - \sin(2 p_{0,1}) p_1 w,
\end{align*}
we introduce their partial derivatives with respect to $p_{0,1}$, $p_1$, and
$w$ as
\begin{align*}
\al_1&=-2p_1w\sin(2p_{0,1})+2w^2\cos(2p_{0,1}),\\
\al_2&=w\cos(2p_{0,1}),\\
\al_3&=p_1\cos(2p_{0,1})+2w\sin(2p_{0,1}),\\
\be_1&=-2w^2\sin(2p_{0,1})-2p_1w\cos(2p_{0,1}),\\
\be_2&=-w\sin(2p_{0,1}),\\
\be_3&=2w\cos(2p_{0,1})-p_1\sin(2p_{0,1}).
\end{align*}
From this we arrive in a straightforward way at the last nonlocal part of
the recursion operator, i.e.,
\[
\begin{bmatrix}
A_1D_x^{-1}\al_2 D_x^{-1}&A_1D_x^{-1}(\al_1D_x^{-1}+\al_3)\\
A_2D_x^{-1}\al_2 D_x^{-1}&A_2D_x^{-1}(\al_1D_x^{-1}+\al_3)
\end{bmatrix}+
\begin{bmatrix}
B_1D_x^{-1}\be_2 D_x^{-1}&B_1D_x^{-1}(\be_1D_x^{-1}+\be_3)\\
B_2D_x^{-1}\be_2 D_x^{-1}&B_2D_x^{-1}(\be_1D_x^{-1}+\be_3)
\end{bmatrix}
\]

So, in the final form we obtain the recursion operator as
\begin{align*}
\CR&=
\begin{bmatrix}
-D_x^2+4u+w^2&-2wD_x^2-w_1D_x+3uw-2w_2\\
2w&-D_x^2+2u+w^2
\end{bmatrix}\\
&+
\begin{bmatrix}
(2u_1+ww_1)D_x^{-1}&(2p_1uw-p_1w_2+2uw_1+3u_1w+2w^2w_1-w_3)D_x^{-1}\\
w_1D_x^{-1}&(-p_1u+u_1+ww_1)D_x^{-1}
\end{bmatrix}\\
&+
\begin{bmatrix}
A_1D_x^{-1}\al_2 D_x^{-1}&A_1D_x^{-1}(\al_1D_x^{-1}+\al_3)\\
A_2D_x^{-1}\al_2 D_x^{-1}&A_2D_x^{-1}(\al_1D_x^{-1}+\al_3)
\end{bmatrix}\\
&+
\begin{bmatrix}
B_1D_x^{-1}\be_2 D_x^{-1}&B_1D_x^{-1}(\be_1D_x^{-1}+\be_3)\\
B_2D_x^{-1}\be_2 D_x^{-1}&B_2D_x^{-1}(\be_1D_x^{-1}+\be_3)
\end{bmatrix}.
\end{align*}

\section{Conclusion}
We gave an outline of the theory of deformations of the equation structure of
differential equations, leading to the construction of recursion operators
for symmetries of such equations. The extension of this theory to the
nonlocal setting of differential equations is essential for getting
nontrivial results. The theory has been applied to the construction of the
recursion operator for symmetries for a coupled KdV--mKdV system, leading to
a highly nonlocal result for this system. Moreover the appearance of
nonpolynomial nonlocal terms in all results, e.g., conservation laws,
symmetries and recursion operator is striking and reveals some unknown and
intriguing underlying structure of the equations.

\section*{Appendix}
Here we present explicit formulas for  $(p_{3,1})_t$,
$(p_{3,2})_t$, $(p_{4,1})_t$, and $(p_5)_t$:
\begin{align*}
(p_{3,1})_t& =(\cos(2 p_{0,1}) (3 p_1^3 u w + p_1^3 w^3 - p_1^3 w_2
+3 p_1^2 u w_1 - 3 p_1^2 u_1 w - 3 p_1^2 w^2 w_1  \\
&- 24 p_1 u^2 w + 105 p_1 u w^3 + 6 p_1 u w_2 - 6 p_1 u_1 w_1
+ 6 p_1u_2 w + 39 p_1 w^5 \\
& - 27 p_1 w^2 w_2 - 96 p_{1,1} p_{1,2} u w^2 - 24 p_{1,1} p_{1,2} w^4
+48 p_{1,1} p_{1,2} w w_2  \\
&- 24 p_{1,1} p_{1,2} w_1^2 + 36 p_3 u w + 12 p_3 w^3 - 12 p_3 w_2 -
12 u^2 w_1  \\
&+ 48 u u_1 w + 39 u w^2 w_1 + 3 u_1w^3 - 6 u_1 w_2 + 6 u_2 w_1 \\
& - 6 u_3 w - 9 w^4 w_1 - 12 w^2 w_3 - 18 ww_1 w_2 + 6 w_1^3)   \\
&+ 2 \sin(2 p_{0,1}) (36 p_1 p_{1,1} p_{1,2} u w + 12 p_1 p_{1,1}p_{1,2}
w^3 - 12 p_1 p_{1,1} p_{1,2} w_2   \\
&+ 54 p_1 u w w_1+ 54 p_1 u_1w^2 + 54 p_1 w^3 w_1
 - 18 p_1 w w_3 + 12 p_{1,1} p_{1,2} u w_1\\
& - 12 p_{1,1} p_{1,2} u_1 w - 12 p_{1,1} p_{1,2} w^2 w_1 - 9 u^2 w^2 + 18 u w^4 + 27 u w w_2
- 9 u w_1^2 \\
&+ 9 u_1 w w_1  + 3 u_2 w^2 + 2 w^6 - 11 w^3 w_2 + 12 w^2w_1^2 - 3 w w_4 + 3 w_1 w_3
- 3 w_2^2) \\
& + 6 p_{1,2} ( - 3 p_1^2 u w - p_1^2 w^3 + p_1^2 w_2 - 2 p_1 u w_1 +
2 p_1 u_1 w + 2 p_1 w^2 w_1  \\
&+ 8 u^2 w + u w^3 - 2 u w_2 + 2 u_1 w_1 - 2 u_2 w - w^5 - 3 w^2 w_2))/12, \\
(p_{3,2})_t& = (2 \cos(2 p_{0,1}) (36 p_1 p_{1,1} p_{1,2} u w
+ 12 p_1 p_{1,1} p_{1,2} w^3 - 12 p_1 p_{1,1} p_{1,2} w_2 \\
& - 54 p_1 u w w_1 - 54 p_1 u_1 w^2 - 54 p_1 w^3 w_1 + 18 p_1 w w_3 +
12 p_{1,1} p_{1,2} u w_1 \\
& - 12 p_{1,1} p_{1,2} u_1 w
- 12 p_{1,1} p_{1,2} w^2 w_1 + 9 u^2 w^2 - 18 u w^4 - 27 u w w_2 + 9u
w_1^2  \\
&- 9 u_1 w w_1
- 3 u_2 w^2 - 2 w^6 + 11 w^3 w_2 - 12 w^2 w_1^2 + 3 w w_4 - 3 w_1w_3
+ 3 w_2^2) \\
& + \sin(2 p_{0,1}) (3 p_1^3 u w + p_1^3 w^3
- p_1^3 w_2 + 3 p_1^2 u w_1 - 3 p_1^2 u_1 w - 3 p_1^2 w^2 w_1  \\
&- 24 p_1 u^2 w + 105 p_1 u w^3 + 6 p_1 u w_2 - 6 p_1 u_1 w_1
+ 6 p_1u_2 w + 39 p_1 w^5 \\
& - 27 p_1 w^2 w_2 + 96 p_{1,1} p_{1,2} u w^2 + 24 p_{1,1} p_{1,2}
w^4- 48 p_{1,1} p_{1,2} w w_2  \\
& + 24 p_{1,1} p_{1,2} w_1^2+ 36 p_3 u w
 + 12 p_3 w^3 - 12 p_3 w_2 - 12 u^2 w_1 + 48 u u_1 w \\
&+ 39 u w^2 w_1  + 3 u_1 w^3 - 6 u_1 w_2
+ 6 u_2 w_1 - 6 u_3 w - 9 w^4 w_1 - 12 w^2 w_3 \\
&- 18 w w_1 w_2 + 6 w_1^3)
+ 6 p_{1,1} ( - 3 p_1^2 u w - p_1^2 w^3 + p_1^2 w_2
- 2 p_1 u w_1+2 p_1 u_1 w  \\
&+ 2 p_1 w^2 w_1 + 8 u^2 w + u w^3 - 2 u w_2 + 2 u_1 w_1
- 2 u_2 w - w^5 - 3 w^2 w_2))/12, \\
(p_{4,1})_t& = (8 \cos(2 p_{0,1}) (3 p_1^3 p_{1,2} u w + p_1^3 p_{1,2} w^3
- p_1^3 p_{1,2} w_2 + 3 p_1^2 p_{1,2} u w_1 \\
& - 3 p_1^2 p_{1,2} u_1 w - 3 p_1^2 p_{1,2} w^2 w_1 + 36 p_1 p_{1,1}^2
p_{1,2} u w + 12 p_1 p_{1,1}^2 p_{1,2} w^3 \\
& - 12 p_1 p_{1,1}^2 p_{1,2}
w_2 - 24 p_1 p_{1,2} u^2 w
 - 3 p_1 p_{1,2} u w^3 + 6 p_1 p_{1,2} u w_2
- 6 p_1 p_{1,2} u_1 w_1  \\
&+ 6 p_1 p_{1,2} u_2 w + 3 p_1 p_{1,2} w^5 + 9
p_1 p_{1,2} w^2 w_2
 + 12 p_{1,1}^2 p_{1,2} u w_1 - 12 p_{1,1}^2p_{1,2} u_1 w \\
& - 12 p_{1,1}^2 p_{1,2} w^2 w_1 - 48 p_{1,1} p_{1,2}^2 u
w^2 - 12 p_{1,1} p_{1,2}^2 w^4 + 24 p_{1,1} p_{1,2}^2 w w_2  \\
&- 12p_{1,1} p_{1,2}^2 w_1^2 + 72 p_{1,1} u^2 w^2 + 18 p_{1,1} u w^4
- 54p_{1,1} u w w_2 + 18 p_{1,1} u w_1^2  \\
&- 18 p_{1,1} u_1 w w_1 - 24p_{1,1} u_2 w^2 - 4 p_{1,1} w^6 - 32 p_{1,1} w^3 w_2 - 24 p_{1,1} w^2
w_1^2  \\
&+ 6 p_{1,1} w w_4- 6 p_{1,1} w_1 w_3  + 6 p_{1,1} w_2^2 + 36
p_{1,2} p_3 u w + 12 p_{1,2} p_3 w^3\\
& - 12 p_{1,2} p_3 w_2 - 12 p_{1,2}u^2 w_1 + 48 p_{1,2} u u_1 w+ 39 p_{1,2} u w^2 w_1
+ 3 p_{1,2} u_1w^3 \\
&- 6 p_{1,2} u_1 w_2 + 6 p_{1,2} u_2 w_1  - 6 p_{1,2} u_3 w - 9p_{1,2} w^4 w_1 - 12 p_{1,2} w^2 w_3\\
&- 18 p_{1,2} w w_1 w_2 + 6p_{1,2} w_1^3)\\
&+ 8 \sin(2 p_{0,1}) (3 p_1^3 p_{1,1} u w + p_1^3 p_{1,1} w^3
- p_1^3 p_{1,1} w_2 + 3 p_1
^2 p_{1,1} u w_1 - 3 p_1^2 p_{1,1} u_1 w  \\
&- 3 p_1^2 p_{1,1} w^2 w_1 + 36 p_1 p_{1,1}
p_{1,2}^2 u w + 12 p_1 p_{1,1} p_{1,2}^2 w^3 - 12 p_1 p_{1,1} p_{1,2}^2 w_2\\
&- 24 p_1 p_{1,1}u^2 w - 3 p_1 p_{1,1} u w^3  + 6 p_1 p_{1,1} u w_2
- 6 p_1 p_{1,1} u_1 w_1 + 6 p_1p_{1,1} u_2 w \\
&+ 3 p_1 p_{1,1} w^5 + 9 p_1 p_{1,1} w^2 w_2 + 48p_{1,1}^2 p_{1,2} u w^2
+ 12 p_{1,1}^2 p_{1,2} w^4 - 24 p_{1,1}^2 p_{1,2} w w_2\\
&+ 12 p_{1,1}^2 p_{1,2} w_1^2 + 12p_{1,1} p_{1,2}^2 u w_1
- 12 p_{1,1} p_{1,2}^2 u_1 w - 12 p_{1,1}p_{1,2}^2 w^2 w_1 \\
& + 36p_{1,1} p_3 u w + 12 p_{1,1} p_3 w^3 - 12 p_{1,1} p_3 w_2
- 12 p_{1,1} u^2 w_1 + 48 p_{1,1}
u u_1 w \\
&+ 39 p_{1,1} u w^2 w_1  + 3 p_{1,1} u_1 w^3
- 6 p_{1,1} u_1 w_2 + 6 p_{1,1} u_2
w_1 - 6 p_{1,1} u_3 w \\
&- 9 p_{1,1} w^4 w_1 - 12 p_{1,1} w^2 w_3  - 18 p_{1,1} w w_1 w_2 +
6 p_{1,1} w_1^3 - 72 p_{1,2} u^2 w^2 \\
&- 18 p_{1,2} u w^4
+ 54 p_{1,2} u w w_2 - 18p_{1,2} u w_1^2  + 18 p_{1,2} u_1 w w_1 + 24 p_{1,2} u_2 w^2 \\
&+ 4 p_{1,2} w^6 + 32 p_{1,2}
w^3 w_2 + 24 p_{1,2} w^2 w_1^2 - 6 p_{1,2} w w_4 \\
& + 6 p_{1,2} w_1 w_3 - 6 p_{1,2} w_2^2) - 3 p_1^4 u w
- p_1^4 w^3 + p_1^4 w_2 - 4 p_1^3 u w_1 + 4 p_1^3 u_1w \\
& + 4 p_1^3 w^2 w_1 - 72 p_1^2 p_{1,1}^2 u w - 24 p_1^2 p_{1,1}^2 w^3 + 24
p_1^2 p_{1,1}^2 w_2 - 72 p_1^2 p_{1,2}^2 u w \\
& - 24 p_1^2 p_{1,2}^2 w^3 + 24 p_1^2
 p_{1,2}^2 w_2 + 48 p_1^2 u^2 w + 6 p_1^2 u w^3 - 12 p_1^2 u w_2 + 12 p_1^
2 u_1 w_1 \\
& - 12 p_1^2 u_2 w - 6 p_1^2 w^5 - 18 p_1^2 w^2 w_2 - 48 p_1 p_{1,1}^
2 u w_1 + 48 p_1 p_{1,1}^2 u_1 w \\
&+ 48 p_1 p_{1,1}^2 w^2 w_1  - 48 p_1 p_{1,2}^2 u w_1 +
 48 p_1 p_{1,2}^2 u_1 w + 48 p_1 p_{1,2}^2 w^2 w_1 - 144 p_1 p_3 u w\\
&- 48 p_1 p_3 w^3 + 48 p_1 p_3 w_2  + 48 p_1 u^2 w_1 - 192 p_1 u u_1 w
- 156 p_1 u w^2 w_1  \\
&-12 p_1 u_1 w^3 + 24 p_1 u_1 w_2 - 24 p_1 u_2 w_1+ 24 p_1 u_3 w + 36 p_1 w^4 w_1
+ 48 p_1 w^2 w_3 \\
&+ 72 p_1 w w_1 w_2  - 24 p_1 w_1^3 + 192 p_{1,1}^2 u^2w + 24p_{1,1}^2 u w^3
- 48 p_{1,1}^2 u w_2 \\
&+ 48 p_{1,1}^2 u_1 w_1- 48 p_{1,1}^2 u_2 w - 24p_{1,1}^2 w^5 - 72 p_{1,1}^2 w^2 w_2  + 192 p_{1,2}^2 u^2 w \\
&+ 24 p_{1,2}^2 u w^3- 48 p_{1,2}^2 u w_2 + 48 p_{1,2}^2 u_1 w_1 - 48 p_{1,2}^2 u_2 w - 24
p_{1,2}^2 w^5 \\
& -72 p_{1,2}^2 w^2 w_2 - 48 p_3 u w_1 + 48 p_3 u_1 w + 48 p_3 w^2 w_1
- 252 u^3w - 36 u^2 w^3 \\
& + 60 u^2 w_2 - 144 u u_1 w_1 + 240 u u_2 w - 7 u w^5 +
342 u w^2 w_2 + 24 u w w_1^2  \\
&+ 144 u_1^2 w + 228 u_1 w^2 w_1 + 94 u_2 w^
3 - 24 u_2 w_2 + 24 u_3 w_1 - 24 u_4 w \\
& - 13 w^7 + 47 w^4 w_2 - 4 w^3 w_1^2
 - 78 w^2 w_4 - 60 w w_1 w_3 \\
&- 120 w w_2^2 + 60 w_1^2 w_2)/48, \\
(p_5)_t& = (54 u^4 + 180 u^3 w^2 - 72 u^2 u_2 + 126 u^2 w^4 - 282 u^2 w w_2
-12 u^2 w_1^2  \\
&- 84 u u_1 w w_1 - 174 u u_2 w^2 + 6 u u_4 + 18 u w^6 - 300
u w^3 w_2 - 90 u w^2 w_1^2  \\
&+ 66 u w w_4 + 6 u w_1 w_3 + 48 u w_2^2 + 42 u_1
^2 w^2 - 6 u_1 u_3 + 12 u_1 w^3 w_1 \\
& + 42 u_1 w w_3 - 48 u_1 w_1 w_2 + 6 u_2^2
- 39 u_2 w^4 + 162 u_2 w w_2 - 48 u_2 w_1^2 \\
& + 48 u_3 w w_1 + 21 u_4 w^2 - 42
w^5 w_2 - 12 w^4 w_1^2 + 35 w^3 w_4 + 120 w^2 w_1 w_3  \\
&+ 195 w^2 w_2^2 +
 120 w w_1^2 w_2 - 6 w w_6 - 30 w_1^4 + 6 w_1 w_5 - 6 w_2 w_4)/6.
\end{align*}

\end{document}